\documentclass{ws-p8-50x6-00} 
\usepackage{graphicx}
\usepackage{latexsym}
  
\begin{document} 

\title{On the Evolution of Nonlinear N-body systems: Substructure of a Distant Galaxy Cluster}

\author{V.G.Gurzadyan (1) and A.Mazure (2)} 

\address{(1) ICRA, Department of Physics, University of Rome "La Sapienza",
Rome, Italy 
and 
Yerevan Physics Institute, Yerevan,
Armenia;
(2) Laboratoire d'Astrophysique de Marseille, Marseille, France
}

\maketitle 

\abstracts{The graph-theoretical S-tree technique developed for the study of the nonlinear N-body systems,
is used for the study for the first time of the substructure of a high redshift cluster of galaxies, namely of MS 1054-03, $z\simeq 0.8$. 
Nearby galaxy clusters studied via the same S-tree code reveal several clearly distinguished subgroups with significant bulk velocities. Therefore the study of distant clusters can enable to trace the evolution
of their dynamics over cosmological time scales. 
The study of the MS 1054-03 showed a different structure, i.e. no evidence of merging subgroups but only of a core of galaxies with velocity dispersion
about 700 $km \cdot s^{-1}$ while the velocity dispersion of the main cluster 1870 $km \cdot s^{-1}$. Further study of samples of high redshift clusters will enable to conclude whether this discrepancy in the structure and dynamics between nearby and distant galaxy clusters is 
a genuine evolutionary effect.}

\section{Introduction}

Fermi-Pasta-Ulam problem has essentially influenced the study of nonlinear N-body systems. Gravitating N-body systems, on one hand, are among systems with complicated properties, on the other hand, globular clusters, galaxies and galaxy clusters are providing possibility to trace observationally many of the obtained conclusions on their dynamics and structure. Particularly, clusters of galaxies which occasionally reveal subgrouping of galaxies in projection, pose the problem of the reconstruction of their 3D-substructure and internal dynamics.

Recent observations supply enough information, including spectroscopical, on high redshift galaxy clusters and therefore provide means to inquire into the evolution of the subclustering properties. Below we report the study of the substructure of the cluster MS1054-03 of redshift 0.8. The analysis is performed by S-tree diagram technique developed for the study of correlations in many dimensional nonlinear systems, in this case of correlations between the parameters of particles of N-body system,  by means of geometrical methods of theory of dynamical 
systems\cite{arn,katok}.

\section{S-tree Graphs and Bounded Subsystems}

Let us mention the main concepts of the S-tree method\cite{book}. Consider a set of objects $\cal A$ which has to be split into sets. If $\cal B(A)$ is the set of all subsets, then the set of all pairs of subsets $\cal (M,N)$ so that $\cal M$ contains $\cal N$, will be
$$
S(\cal A)=\bigcup_{\cal M \in \cal B(\cal A)} \bigcup_{N \in B(M)} (\cal M,\cal N) \subset B(A) \times \cal B(\cal A).
$$
One can introduce a non-negative function $\cal P$, the boundness function to describe the degree of the boundedness of $\cal N$ in $\cal M$
$$
\cal P : S(\cal A) \rightarrow R_+ : \cal (M,N) \rightarrow P_M(N).
$$ 
Various choices for the function $P$ to correspond the correlation with respect the subgrouping are discussed in\cite{book}, for example, the choice of the projected distances corresponds to the two-point correlation function. Theory of dynamical systems provides a detailed characteristic for N-dimensional systems, the two-dimensional curvature of the configurational space
$$
K_{\mu\nu}=R_{\mu\nu\lambda\rho}u^{\lambda}u^{\rho},
$$
where $R_{\mu\nu\lambda\rho}$ is the Riemann tensor, $u^{\lambda}$ is the velocity of geodesics of the phase space, $\lambda = (a,i), a=1,...,N;
i=1,2,3$.
The two dimensional curvature is a known tool also for the study of the instability of the many-dimensional systems\cite{arn}. Therefore its use enables to link the statistical, chaotic properties of the systems with the emergence of the substructures and their properties.
The versions of S-tree used in the studies of clusters of galaxies are applying the two-dimensional curvature for the function $P$. S-tree technique is developed further in\cite{Bek}.
Another application of similar graph-theoretical methods is the possibility of comparison of the results of numerical simulations with observed large-scale galaxy distributions\cite{ueda}.

\section{Galaxy Associations}

S-tree method has been applied to a number of cluster of galaxies, including Coma and Virgo\cite{GM,GM01}. Various subgroups of galaxies have been detected, with indications for their essential bulk motions.
While studying the substructure of Abell clusters of ENACS\cite{mazure} (European Nearby Abell Cluster Survey of ESO), this method enabled to reveal subgroups of galaxies, denoted as galaxy associations, with marked dynamical properties\cite{GM}. The studied sample of ENACS clusters with redshifts up to  $z \simeq 0.1$, is shown to contain typically 2-3 subgroups with velocity dispersion around 100-200 $km\cdot s^{-1}$ with truncated Gaussian velocity distribution. The latter was interpreted as stripping of the subgroup while moving within the cluster. In the case of Coma cluster we had detected a subgroup which has practically dissolved within the main cluster, while another one with survived dynamical identity. Thus, the substructure of nearby clusters supports the view of merging of several subgroups, and hence, non-stationary processes ongoing within the clusters.

That remarkable conclusion had to indicate the primordial nature of the subgroups, galaxy associations, and their emergence with essential bulk velocities. Various consequences of these predictions can be checked via observations, namely, the different star forming properties of the galaxies of the associations and of the main cluster, non-identical properties of the spiral galaxies, of the populations of galactic and intracluster globular clusters. The latter effect has been studied in\cite{mario} 
for the subgroup 2 of the Coma cluster detected in\cite{GM01}.

In which extent this picture can be accommodated within the cosmological evolution of the clusters? The answer to this question can be anticipated only via the study of high redshift galaxy clusters.

\begin{table}
\caption{The parameters of the galaxy cluster MS 1054-03 and of its cores}
\begin{tabular}{l|c|c|c|c|c}
\hline
 System                                & Main cluster & core-1  & core-2\\
\hline 
Number of galaxies                     & 68           & 35      & 10\\
\hline
Median velocity, $km \cdot s^{-1}$     & 249,148      & 249,903 & 249,651\\
\hline
Velocity dispersion, $km \cdot s^{-1}$ & 1872        & 775     & 708 \\
\hline
\end{tabular}
\end{table}

\section{MS 1054-03 Substructure}
  
The dataset\cite{vanD} that we used for the galaxy cluster MS 1054-03, contained 73 galaxies with redshifts between $z=0.81320 - 0.84700$ and stellar magnitudes in the range $m_B = 19.530-23.210$.
That dataset has been studied by the very same S-tree code used for the study of nearby clusters, i.e. we deal with identical procedures and computational requirements.
The results of the analysis of the substructure of MS 1054-03 are given in Fig.1 of the redshift histograms of the galaxies and Table 1. The main cluster according to S-tree appeared to contain 68 galaxies (of 73 in the initial catalog), of velocity dispersion 1872 
$km\cdot  s^{-1}$. The internal substructure of that system revealed only one main system of 35 galaxies (core-1), of velocity dispersion 775 $km \cdot s^{-1}$. Its more strongly bounded core (core-2) contains 10 galaxies of velocity dispersion
708 $km \cdot s^{-1}$. Note, although while moving from core-1 to core-2 the number of galaxies decreases drastically, the velocity dispersion does not decrease much. This indicates that while a number of galaxies of lower mass (higher stellar magnitude) are absent in the catalog, nevertheless the main galaxies taken into account at the calculations reflect for the genuine dynamics of the core.  
\begin{table}
\caption{The galaxies of the core-2 of MS 1054-03}
\begin{tabular}{l|c|c|c|c|c|c|c}
\hline
ID2     &X          &Y       & Z  & BTz   &MORPH\\ 
\hline 
696    &-58.080   &-83.6900  &0.83170   &21.720  &S0/a\\    
703    &-24.530   &-79.3300  &0.83230   &21.910  &?\\       
777    & 90.590   &-63.5400  &0.83230   &22.160  &S0\\     
1114   & -3.626   &-31.4200  &0.83170   &22.610  &S0/a\\    
1198   &-58.120   &-25.0600  &0.83130   &23.360  &Sa\\    
1325   &52.780    &-9.7500   &0.83110   &21.050  &E\\      
1484   & 0.000    &0.0000    &0.83180   &20.700  &E\\       
1520   &-21.680   &19.1700   &0.83100   &22.740  &E/S0\\    
1585   &-53.000   &27.8600   &0.83690   &23.280  &Sc\\     
1758   &68.440    &55.7000   &0.83740   &23.110  &S0/a\\    
\hline
\end{tabular}
\end{table}
The apparent non-Gaussian velocity distribution of the cluster galaxies might be either intrinsic or due to the non-completeness of the sample, or both. 
The core-2 of 10 galaxies is dominated by E and S0 types, namely,
it contains 3 S0/a, S0, 2 E and E/S0 galaxies (Table 2). Note, that $z\simeq 0.5$
clusters show increased fraction of spirals by a factor of 2-3, increase of fraction
of ellipticals but decrease those of S0 galaxies\cite{Dressler}.

Thus, MS 1054-03 revealed a different dynamical substructure as compared with the nearby Abell clusters, containing a single bounded core rather than several subgroups as the latter ones. To understand whether the absence of merging subgroups is a trend
of cosmological evolution will need further study of distant clusters.

\begin{figure}[htp]
\begin{center}
\includegraphics[width=\hsize,clip]{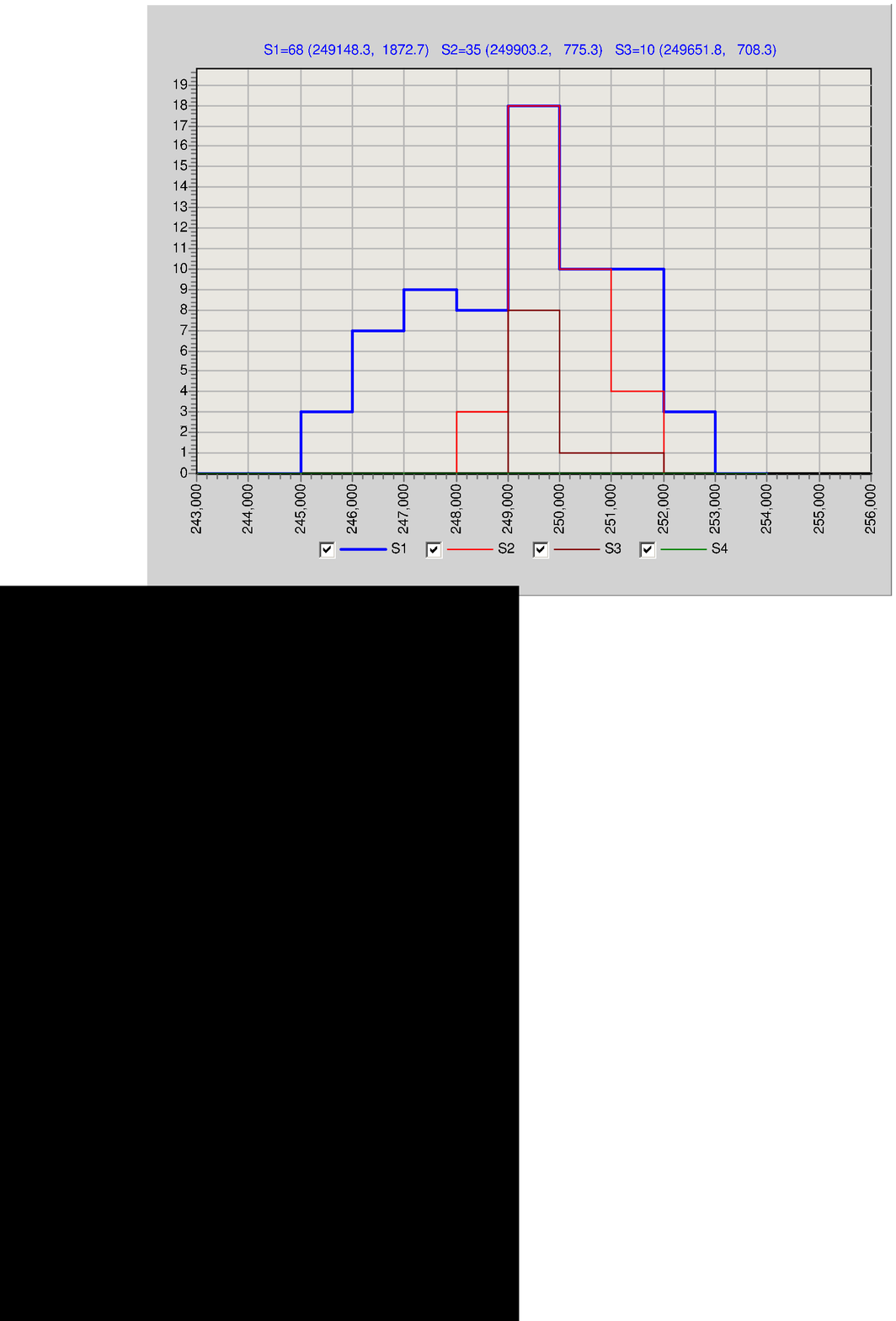}
\caption{The histogram of the redshifts of the cluster of MS1054-03. The main system (S1) contains 68 galaxies, 
its more strongly bounded core-subsystem is shown at two degrees of boundness, i.e. as 
the core-1 (S2) containing the core-2 (S3), with 35 and 10 galaxies, respectively.}
\end{center}
\end{figure}

We thank C.Adami for assistance with the dataset and A.Melkonian with the S-tree code. 
VG's visit to Marseille was supported by French-Armenian Jumelage.


\end{document}